Pre-inflation models and WMAP data

Shiro Hirai *

Department of Digital Games, Faculty of Information Science and Arts,
Osaka Electro-Communication University
1130-70 Kiyotaki, Shijonawate, Osaka 575-0063, Japan
*E-mail*: hirai@isc.osakac.ac.jp

**Abstract**

The effect of pre-inflation physics on the power spectrum of scalar perturbations is estimated and there is a possibility that the pre-inflation physics explains the Wilkinson Microwave Anisotropy Probe data if the length of inflation is near 60 *e*-folds. Considering various pre-inflation models with radiation-dominated or matter-dominated periods before inflation, the power spectrum of curvature perturbations for large scales is calculated, and the spectral index and running spectral index are derived.

**1. Introduction**

Recently released data from the Wilkinson Microwave Anisotropy Probe (WMAP) has brought surprising results for astrophysics, and has allowed a number of cosmological parameters to be fixed precisely. This study focuses on the following four results [1]: (1) the spectral index varies from $n_s > 1$ on large scales to $n_s < 1$ on small scales, (2) $n_s = 1.10$ on the scale $k_0 = 0.002$ Mpc$^{-1}$, (3) $dn_s/d\ln k = -0.042$ on the scale $k_0 = 0.002$ Mpc$^{-1}$, and (4) suppression of the spectrum at large angular scales.

Many attempts to explain the data obtained by the WMAP have been proposed. As simple slow-roll inflation models appear to be unable to explain these four results adequately [2], double inflation models and more complicated models have been considered [3]. The suppression of the spectrum at large angular scales has been suggested to be representative of a finite-size Universe with nontrivial topology or a closed Universe [4]. Others have suggested that these four results imply a new physics [5], although it has been subsequently pointed out that the $k$-dependence of the new physics does not follow directly and that the contribution is weak [6]. In this study, these four results are suggested to represent the contribution of pre-inflation physics rather than a new physics.

Recently, we considered the effect of the initial condition in inflation on the power spectrum of curvature perturbations [7]. Based on the physical conditions before inflation, the possibility exists that the initial state of scalar perturbations in inflation is not simply the Bunch-Davies state, but also a more general state – a squeezed state, as occurs between two phases, such as between the inflation and radiation-dominated epochs [8]. A formula for the power spectrum of curvature perturbations having any initial conditions in inflation was obtained, consisting of the familiar formation multiplied by a factor indicating the contribution of an initial condition. In this paper, formulae for the spectral index and the running spectral index are derived for any initial conditions of inflation, and the physical meanings of the derived formulae are clarified through examination of a number of pre-inflationary cosmological models. The matching conditions for the scalar perturbations, which appear to be essential and which have been discussed in relation to inflation [9] and an ekpyrotic scenario [10] are also examined.

Calculations are made assuming two matching conditions; one in which the gauge potential and its first $\eta$-derivative are continuous at the transition point, and one in which the transition happens on a hyper-surface of constant energy as proposed by Deruelle and Mukhanov [9]. The differences between the two matching conditions are investigated in detail by calculating the power spectrum of curvature perturbations (i.e., the factor expressing the effect of pre-inflation physics), the spectral index, and the running spectral index for two models and two cases. The derived factors indicating the contribution of pre-inflation physics have many interesting features, such as small $k$-dependence on large scales, and a dependence on the start time of inflation. The value of these pre-inflation modes as an explanation for the WMAP data is also discussed.

This paper is organized as follows. In section 2, a formula for the power spectrum of curvature perturbations is re-derived for any initial conditions in inflation. In section 3, the spectral index and running spectral index are discussed. In section 4, some pre-inflation models are considered In section 5, using the derived formula, the effect of pre-inflation models on the power spectrum, the spectral index and the running spectral index are estimated taking matching conditions into consideration, and the results are compared with WMAP data. In section 6, the results obtained in the present study are discussed at length.

## 2. Scalar perturbations

The formula for the power spectrum of curvature perturbations in inflation is derived here for any initial conditions by applying a commonly used method [11]. This formula was originally derived in [12, 7], but as it represents a critical result, it is derived again here. As a background spectrum, we consider a spatially flat Friedman-Robertson-Walker (FRW) universe described by metric perturbations. The line element for the background and scalar metric perturbations is generally expressed as [13]

$$ds^2 = a^2(\eta)\{(1+2A)d\eta^2 - 2\partial_i B dx^i d\eta - [(1-2\Psi)\delta_{ij} + 2\partial_i\partial_j E]dx^i dx^j\} \quad (1)$$

where is $\eta$ the conformal time. The density perturbation in terms of the intrinsic curvature perturbation of comoving hypersurfaces is given by

$$\boldsymbol{R} = -\Psi - \frac{H}{\dot{\phi}}\delta\phi, \quad (2)$$

where $\phi$ is the inflaton field, $\delta\phi$ is the fluctuation of the inflaton field, $H$ is the Hubble expansion parameter, and $\boldsymbol{R}$ is the curvature perturbation. Overdots represent derivatives with respect to $t$, and the prime represents the derivative with respect to the conformal time $\eta$. If the gauge-invariant potential $u \equiv a(\delta\phi + \frac{\dot{\phi}}{H}\Psi)$ is introduced, the action (Lagrangian) for scalar perturbations is written as [14]

$$S = \int d^4 x L$$

$$= \frac{1}{2}\int d\eta d^3 x \{(\frac{\partial u}{\partial \eta})^2 - c_s^2 (\nabla u)^2 + \frac{Z''}{Z} u^2\}, \quad (3)$$

where $c_s$ is the velocity of sound, $Z = \frac{a\dot{\phi}}{H}$, and $u = -Z\boldsymbol{R}$. The field $u(\eta, \boldsymbol{x})$ is expressed using annihilation and creation operators as follows.

$$u(\eta, x) = \frac{1}{(2\pi)^{3/2}} \int d^3k \{ a_k \, u_k(\eta) + a_{-k}^{\dagger} u^*{}_k(\eta) \} e^{-ikx}. \tag{4}$$

The field equation of $u_k$ is derived as

$$\frac{d^2 u_k}{d\eta^2} + (c_s^2 k^2 - \frac{1}{Z}\frac{d^2 Z}{d\eta^2}) \, u_k = 0. \tag{5}$$

The solution $u_k$ satisfies the normalization condition $(u_k du_k^*/d\eta) - (u_k^* du_k/d\eta) = i$. We consider a power-law inflation $a(\eta) \approx (-\eta)^p \, (= t^{p/(p+1)})$, in which case equation (5) is written as

$$\frac{d^2 u_k}{d\eta^2} + (k^2 - \frac{p(p-1)}{\eta^2}) u_k = 0, \tag{6}$$

where $c_s^2 = 1$ in the scalar field case. The solution of equation (6) is written as

$$f_k^I(\eta) = i\frac{\sqrt{\pi}}{2} e^{-ip\pi/2} (-\eta)^{1/2} H_{-p+1/2}^{(1)}(-k\eta), \tag{7}$$

where $H_{-p+1/2}^{(1)}$ is the Hankel function of the first kind of order $-p + 1/2$. As a general initial condition, the mode function $u_k(\eta)$ is assumed to be

$$u_k(\eta) = c_1 \, f_k^I(\eta) + c_2 \, f_k^{I*}(\eta), \tag{8}$$

where the coefficients $c_1$ and $c_2$ obey the relation $|c_1|^2 - |c_2|^2 = 1$. The important point here is that the coefficients $c_1$ and $c_2$ do not change during inflation. In ordinary cases, the field $u_k(\eta)$ is considered to be in the Bunch-Davies state, i.e., $c_1 = 1$ and $c_2 = 0$, because as $\eta \to -\infty$, the field $u_k(\eta)$ must approach plane waves, e.g., $e^{-ik\eta}/\sqrt{2k}$.

Next, the power spectrum is defined as follows [11].

$$<R_k(\eta), R_l^*(\eta)> = \frac{2\pi^2}{k^3} P_R \delta^3(k-l), \tag{9}$$

where $R_k(\eta)$ is the Fourier series of the curvature perturbation $R$. The power spectrum $P_R^{1/2}$ is then written as follows [11].

$$P_R^{1/2} = \sqrt{\frac{k^3}{2\pi^2}} \left|\frac{u_k}{Z}\right|. \tag{10}$$

The power spectrum is calculated using the approximation of the Hankel function. The series $H^{(1)}_{-p+1/2}(z)$ at the limit of $z \to 0$ ($z = -k\eta$) is written as follows [15].

$$H^{(1)}_{-p+1/2}(z) = z^{p-1/2}\left(-\frac{i2^{1/2-p}\sec p\pi}{\Gamma(1/2+p)} + \frac{i2^{-3/2-p}z^2 \sec p\pi}{\Gamma(3/2+p)} - \frac{i2^{-9/2-p}z^4 \sec p\pi}{\Gamma(5/2+p)}\right.$$

$$+ o[z]^6)$$

$$+ z^{-p-1/2}\left(\frac{2^{-1/2+p}(1+i\tan p\pi)z}{\Gamma(3/2-p)} - \frac{i2^{-5/2+p}(-i+\tan p\pi)z^3}{\Gamma(5/2-p)} - \frac{2^{-11/2+p}(1+i\tan p\pi)z^5}{\Gamma(7/2-p)}\right.$$

$$+ o[z]^7), \tag{11}$$

where $\Gamma(-p+1/2)$ expresses the Gamma function. In the present study, the term $z^{p-1/2}$ ($= (-k\eta)^{p-1/2}$) is of leading order. The power spectrum of the leading and next leading correction of $-k\eta$ in the case of squeezed initial states can be written as

$$P_R^{1/2} = (2^{-p}(-p)^p \frac{\Gamma(-p+1/2)}{\Gamma(3/2)} \frac{1}{m_P^2} \frac{H^2}{|H'|})|_{k=aH} (1 - \frac{(-k\eta)^2}{2(1+2p)})$$

$$\times |c_1 e^{-ip\pi/2} + c_2 e^{ip\pi/2}|.$$

$$= (\frac{H^2}{2\pi\dot\phi})|_{k=aH} |c_1 e^{-ip\pi/2} + c_2 e^{ip\pi/2}|. \tag{12}$$

This formula differs slightly from Hwang's formula [12] due to the introduction of the term $e^{-ip\pi/2}$ into equation (7), as required in order that in the limit $\eta \to -\infty$, the field $u_k(\eta)$ must approach plane waves [10]. The quantity $C(k)$ is defined as

$$C(k) = c_1 e^{-ip\pi/2} + c_2 e^{ip\pi/2}. \tag{13}$$

If $|C(k)| = 1$, the leading term of $-k\eta$ in equation (12) can be written as $P_R^{1/2} = (\frac{H^2}{2\pi\dot{\phi}})|_{k=aH}$ [16]. However, equation (12) implies that if, under some physical circumstances, the Bunch-Davies state is not adopted as the initial condition of the field $u_k$, the possibility exists that $|C(k)| \neq 1$. Then, the next problem is whether such a situation actually exists. Here, a number of pre-inflation toy models are assumed, such as a situation in which the pre-inflation is a radiation-dominated period, as is quite probable. A state of $u_k$ is know to become a squeezed state at the initial time of inflation, that is, the state of $u_k$ is not the Bunch-Davies state [7]. In section 5, the value of $|C(k)|$ is calculated using a number of pre-inflation models.

## 3. Spectral index and running spectral index

For the spectral index of curvature perturbations, if it is assumed that $|C(k)|^2 \propto k^{\Delta n}$, then $P_R$ can be written as $P_R \propto k^{n-1+\Delta n}$, where $n - 1$ is the ordinary contribution of the spectral index (i.e., the case $c_1 = 1$ and $c_2 = 0$). However, actual calculation of the spectral index and running spectral index presents some problems. The spectral index $n_s - 1$ and running spectral index $\alpha$ are defined as

$$n_s - 1 \equiv d\log|P_R|/d\log k, \tag{14}$$

$$\alpha \equiv dn_s/d\log k, \tag{15}$$

Another definition using the Taylor series is

$$\log|P_R(k)| \equiv P_R(k_0) + (n_s - 1)\log(\frac{k}{k_0}) + \frac{\alpha}{2}(\log(\frac{k}{k_0}))^2 \tag{16}$$

where $k_0$ is the pivot point. If the spectral index and running spectral index are calculated using equations (14), (15) or (16), special attention is necessary. First, consider the spectral index. For example, consider the power spectrum in the case of $z \to \infty$ ($z = k\eta_2$), which is derived in equation (27) as follows.

$$P_R(k) = \frac{2 + \cos(p\pi + 2z)}{\sqrt{3}} + \frac{p(p-1)\sin(p\pi + 2z)}{\sqrt{3}z}$$

$$+ \frac{(1-p)p - p(-2 + 3p - 2p^2 + p^3)\cos(p\pi + 2z)}{2\sqrt{3}z^2} \ . \tag{17}$$

This means $n_s - 1 \cong 0$. However, using only the leading order of equation (17), the spectral index can be calculated from equation (14) as follows.

$$n_s - 1 \cong \frac{-2z\sin(p\pi + 2z)}{2 + \cos(p\pi + 2z)} \ . \tag{18}$$

This result seems improper, and this situation also occurs in the case of using equation (16). The running spectral index also encounters the same problem. Moreover, if $P_R(k) \propto k^m$ ($m$ is a constant), the running spectral index becomes zero. In order to obtain a non-zero running index, we consider the case $P_R(k) \propto a_1 k^m + a_2 k^{m-1}$ (next leading term), or where $m$ is dependent on $k$.

Thus, the calculations of the spectral index and the running spectral index are quite delicate. Later in this paper, with consideration for this point, $\Delta n = d\log|C(k)|^2/d\log k$, and $\Delta\alpha \equiv d\Delta n/d\log k$, representing the contribution of pre-inflation, are calculated for a range of simplified cosmological models.

Assuming power-law inflation, the spectral index $n_s - 1$ is written as

$$n_s - 1 = -2(p+1)/p + \Delta n, \tag{19}$$

where the term $-2(p+1)/p$ is the effect of the power-law inflation [11]. However, the running spectral index is written as $\alpha = \Delta\alpha$, because the term $-2(p+1)/p$ is a constant.

## 4. Pre-inflationary cosmological models

The effect of pre-inflation physics is examined using simplified models of pre-inflation as an illustration. Here, the pre-inflation model is considered to consist simply of a radiation-dominated period or a scalar-matter-dominated period. A simple cosmological model is assumed, as defined by

$$\text{Pre-inflation} \quad a_P(\eta) = b_1(-\eta - \eta_j)^r,$$

$$\text{Inflation} \quad a_I(\eta) = b_2(-\eta)^p, \tag{20a}$$

where

$$\eta_j = (\frac{r}{p} - 1)\eta_2, \quad b_1 = (\frac{p}{r})^r(-\eta_2)^{p-r} b_2. \tag{20b}$$

The scale factor $a_I(\eta)$ represents ordinary inflation. If $p = -1$, the inflation is de-Sitter inflation, and if $p < -1$, the inflation is a power-law inflation ($p = -10/9$, $a(t) = t^{10}$). Inflation is assumed to begin at $\eta = \eta_2$ and end at $\eta = \eta_3$. In pre-inflation, for the case $r = 1$, the scale factor $a_P(\eta)$ indicates that a radiation-dominated period occurs, whereas for the case $r = 2$, the scale factor $a_P(\eta)$ indicates that a scalar-matter-dominated period occurs. Here, the period of inflation is assumed to be sufficiently long, that is, in the plot of $|C(k)|^2$, $n_s - 1$, and $\alpha$, the start time of inflation $\eta_2$ is chosen as the time at which perturbations of the current Hubble horizon size exceed the Hubble radius in inflation.

## 5. Calculation of power spectrum, spectral index, and running spectral index

Using the pre-inflation models and taking account of the matching conditions, the power spectrum $|C(k)|$ is calculated and used to derived the spectral index and running spectral index. The differences between models and matching conditions are obtained, and the results are compared with the WMAP data.

### 5.1 Radiation-dominated period before inflation

Consider the case of a radiation-dominated period before inflation. In this case, the scale factor becomes $a_P(\eta) = b_1(-\eta - \eta_j)$, i.e., $r = 1$. In the radiation-dominated period, the field equation $u_k$ can be written as equation (5). In this case, $Z$ is written as $Z = a_P (2(\mathbf{H}^2 - (\mathbf{H})')/3)^{1/2} /(c_s \mathbf{H})$ [17,18], where $\mathbf{H} = (a_P)'/a_P$. The value of $c_s^2$ can be fixed at 1/3. The solution of equation (5) is then given by [11]

$$f_k^R(\eta) = \frac{3^{1/4}}{\sqrt{2k}} e^{-ik(\eta+\eta_j)/\sqrt{3}}. \tag{21}$$

Here, for simplicity, it is assumed that the mode function of the radiation-dominated period can be written as equation (21). The solution of equation (6) in inflation becomes equation (7). Although an inflationary model in the presence of gauge fields was considered in [19], the contribution of the radiation field is neglected in the present analysis of inflation for simplicity. The general mode function in inflation can therefore be written as

$$u_k^I(\eta) = c_1 f_k^I(\eta) + c_2 f_k^{I*}(\eta). \tag{22}$$

In order to fix the coefficients $c_1$ and $c_2$, a matching condition in which the mode function and first $\eta$-derivative of the mode function are continuous at the transition time $\eta = \eta_2$ is employed, where $\eta_2$ is the start time of inflation. The coefficients $c_1$ and $c_2$ can then be calculated as follows.

$$c_1 = \frac{\sqrt{\pi}}{2 \cdot 3^{3/4} \sqrt{2z}} e^{i(p\pi/2 + z/\sqrt{3}p)} ((3-3p-i\sqrt{3}z) H_{-p+1/2}^{(2)}(z) - 3z H_{-p+3/2}^{(2)}(z)), \tag{23}$$

$$c_2 = \frac{\sqrt{\pi}}{2 \cdot 3^{3/4} \sqrt{2z}} e^{i(-p\pi/2 + z/\sqrt{3}p)} ((3-3p-i\sqrt{3}z) H_{-p+1/2}^{(1)}(z) - 3z H_{-p+3/2}^{(1)}(z)), \tag{24}$$

where $z = -k\eta_2$. The quantity $C(k)$ is derived from equation (13) as

$C(k) =$

$$-\frac{\sqrt{\pi}}{2 \cdot 3^{3/4} \sqrt{2z}} e^{iz/\sqrt{3}p} \{(-3+3p+i\sqrt{3}z)(H_{-p+1/2}^{(1)}(z) + H_{-p+1/2}^{(2)}(z)) + 3z(H_{-p+3/2}^{(1)}(z) + H_{-p+3/2}^{(2)}(z))\}.$$

$$\tag{25}$$

Since we are interested in the super-large scale, i.e., $z \ll 1$, the Hankel function is expanded around $z = 0$ (equation (11)). In the case of $z \to 0$, the quantity $C(k)$ can be obtained in a simple form as

$$|C(k)|^2 \cong \frac{4^{-1+p}\pi(-9+24p-21p^2+6p^3+z^2(6-10p+3p^2))z^{-2p}}{\sqrt{3}(-3+2p)(\Gamma(\frac{3}{2}-p))^2}. \qquad (26)$$

If we consider the case $p = -10/9$, the quantity $|C(k)|$ is proportional to $z^{10/9}$, when $z \to 0$ ($k \to 0$), and $|C(k)|$ becomes zero.

Next, we consider the case of $z \to \infty$ (small-scale cases), for which the quantity $|C(k)|$ is approximately given by

$$|C(k)|^2 \cong \frac{2+\cos\theta}{\sqrt{3}} + \frac{p(p-1)\sin\theta}{\sqrt{3}z} + \frac{(1-p)p - p(-2+3p-2p^2+p^3)\cos\theta}{2\sqrt{3}z^2}, \qquad (27)$$

where $\theta = p\pi + 2z$. From equation (27), the leading contribution of $|C(k)|^2$ does not depend on $p$, and oscillates around $2/\sqrt{3}$. The numerical value from equation (27) is $0.577 \leq |C(k)|^2 \leq 1.732$.

The quantity $|C(k)|^2$, representing the contribution of pre-inflation ($k$-dependence), is plotted as a function of $z$ ($= -k\eta_2$) in figure 1, where the case $p = -10/9$ is considered. Here, although the value $p = -10/9$ is used, other values (e.g. $p = -100/99$) gives similar results. The graph is plotted such that inflation starts at $\eta_2$ ($= -z/k$), denoting the time at which the perturbations of the current Hubble horizon size exceed the Hubble radius in inflation. That is, the length of inflation is assumed to be 60 $e$-folds. Perturbations of the current Hubble horizon size for other lengths of inflation can be found in this figure. For example, if the length of inflation is 65 *e*-folds, the value of $z = 90$ ($p = -10/9$) gives the perturbations of the current Hubble horizon size. If a longer inflation exists, from the super horizon scales to small scales, the quantity $|C(k)|^2$ behaves according to equation (27), that is, it oscillates around $2/\sqrt{3}$. Although the results for a radiation-dominated period before inflation are not altogether surprising, it is important to note that even a very long inflationary period cannot remove the vibration of $|C(k)|^2$ around $2/\sqrt{3}$.

Next, we consider the spectral index of the contribution of pre-inflation $\Delta n$. From equations (26) and (27), the asymptotic form of $\Delta n$ can be obtained: $\Delta n \to -2p$ for $z \to 0$, and $\Delta n \to 0$ for $z \to \infty$. From equation (19) $z \to 0$, $n_s - 1 \to -2p - 2(p+1)/p$ and $z \to \infty$, $n_s - 1 \to -2(p+1)/p$. For example, when $p = -10/9$, $n_s - 1$ changes from 2.022 to 0.222, from super large scale to small scale, and when $p = -100/99$, $n_s - 1$ changes from 2.000 to 0.020. In detail, in the case of $z \to 0$, using equation (26) the spectral index and the running spectral index can be calculated as follows.

$$\Delta n = \frac{d|C_k(z)|^2}{d \log k}$$

$$\cong -2p + \frac{2(6-10p+3p^2)z^2}{3(-1+p)^2(-3+2p)} + \frac{2p(-66+128p-68p^2+9p^3)z^4}{9(-1+p)^4(3-2p)^2(-5+2p)}, \qquad (28)$$

$$\Delta \alpha = \frac{d\Delta n}{d \log k} \cong \frac{4(6-10p+3p^2)z^2}{3(-1+p)^2(-3+2p)} + \frac{8p(-66+128p-68p^2+9p^3)z^4}{9(-1+p)^4(3-2p)^2(-5+2p)}. \qquad (29)$$

However, in the case of $z \geq 2$, the contribution of the terms $sin(p\pi + 2z)$ and $cos(p\pi + 2z)$ preclude such calculation. For this reason, the spectral index $n_s - 1$ is only shown for the range $0 \leq z \leq 3$ in figure 2, and the running spectral index $\alpha (=\Delta \alpha)$ is plotted for the same range in figure 3.

The behavior of the power spectrum is very interesting, because the value of $|C(k)|^2$ decreases on the super large scale, apparently consistent with the WMAP data (4) assuming 60 $e$-folds for the length of inflation. From figure 2, the spectral index $n_s - 1$ may similarly vary from $n_s > 1$ on a large scale to $n_s < 1$ on a small scale. In some range the running spectral index is negative from figure 3.

**5.2 Scalar-matter-dominated period before inflation**

In the case of a scalar-matter-dominated period before inflation, in which the scalar-matter is the inflaton field $\phi$, the scale factor becomes $a_P = b_1 (-\eta - \eta_j)^2$, i.e., $r = 2$. The solution of equation (6) is this case is given by

$$f_k^S(\eta) = \frac{1}{\sqrt{2k}} (1 - \frac{i}{k(\eta + \eta_j)}) e^{-ik(\eta + \eta_j)}. \tag{30}$$

Here, for simplicity, it is assumed that the mode function of the scalar-dominated period can be written as $f_k^S(\eta)$. The solution of equation (6) in inflation is equation (7), allowing the general mode function in inflation to be written as equation (8). In order to fix the coefficients $c_1$ and $c_2$, the mode function and first $\eta$–derivative of the mode function are assumed to be continuous at $\eta = \eta_2$. The coefficients $c_1$ and $c_2$ can be calculated as follows.

$$c_1 = \frac{-i\sqrt{\pi}}{8\sqrt{2z^3}} e^{i(p\pi/2 + 2z/p)} ((p^2 + 4z(i+z) - 2p(1+iz)) H^{(2)}_{-p+1/2}(z) + 2(p-2iz)z \, H^{(2)}_{-p+3/2}(z)), \tag{31}$$

$$c_2 = \frac{-i\sqrt{\pi}}{8\sqrt{2z^3}} e^{i(-p\pi/2+2z/p)} ((p^2+4z(i+z)-2p(1+iz)) H^{(1)}_{-p+1/2}(z) + 2(p-2iz)z\, H^{(1)}_{-p+3/2}(z)). \quad (32)$$

The quantity $C(k)$ is derived from equation (13) as

$$C(k) = \frac{-i\sqrt{\pi}}{4\sqrt{2z^3}} e^{2iz/p} \{(p^2+4z(i+z)-2p(1+iz))(H^{(1)}_{-p+1/2}(z) + H^{(2)}_{-p+1/2}(z))$$

$$+ 2(p-2iz)z\,(H^{(1)}_{-p+3/2}(z) + H^{(2)}_{-p+3/2}(z))\}. \quad (33)$$

At super large scales, where $z \ll 1$, $|C(k)|^2$ can be obtained using equation (11) in simple form as

$$|C(k)|^2 \cong \frac{4^{-3+2p}\pi(-2+p)^2(-3p^2+2p^3+(-12+20p+p^2)z^2)z^{-2-2p}}{(-3+2p)(\Gamma(\frac{3}{2}-p))^2}. \quad (34)$$

For $p = -10/9$, the coefficients $|C(k)|$ are proportional to $z^{1/9}$ when $z \to 0$ ($k \to 0$), and so $|C(k)|$ becomes zero. For $z \to \infty$ (small scales), $|C(k)|^2$ is obtained as

$$|C(k)|^2 \cong 1 - \frac{p(p-2)\cos(p\pi+2z)}{4z^2}. \quad (35)$$

In this case, $|C(k)|^2 \cong 1$, which is different from the case for the radiation-dominated period before inflation. The quantity $|C(k)|^2$, which shows the contribution of pre-inflation ($k$-dependence), is plotted in figure 4 as a function of $z$ ($=-k\eta_2$) for the case of $p = -10/9$.

Next, we consider the spectral index $\Delta n$, that is, the contribution of pre-inflation. First, from equations (34) and (35), the asymptotic form of $\Delta n$ can be obtained: $\Delta n \to -2-2p$ for $z \to 0$, and $\Delta n \to 0$ for $z \to \infty$. From equation (19), $z \to 0$, $n_s - 1 \to -2 - 2p - 2(p+1)/p$ and $z \to \infty$, $n_s - 1 \to -2(p+1)/p$. For example, when $p = -10/9$, $n_s - 1$ changes from 0.022 to 0.222 from super large scale to small scale, and when $p = -100/99$, $n_s - 1$ changes from 0.0002 to 0.020. In detail, in the case of $z \to 0$, using equation (34), the spectral index and running spectral index can be calculated as follows.

$$\Delta n \cong -2(1+p) + \frac{2(-12+20p+p^2)z^2}{p^2(-3+2p)} + \frac{2(720-2688p+2840p^2-792p^3+p^4)z^4}{p^4(3-2p)^2(-5+2p)},$$

(36)

$$\Delta \alpha \cong \frac{4(-12+20p+p^2)z^2}{p^2(-3+2p)} + \frac{8(720-2688p+2840p^2-792p^3+p^4)z^4}{p^4(3-2p)^2(-5+2p)}.$$
(37)

However, in the case of $z \geq 2$, the same as in the radiation-dominated case, there are no valid methods for the calculation due to the contribution of the terms $sin(p\pi + 2z)$ and $cos(p\pi + 2z)$. Therefore, $n_s - 1$ and $\alpha(=\Delta\alpha)$ are only given for the range $0 \leq z \leq 3$ in figures 5 and 6.

The behavior of $|C(k)|^2$, $n_s - 1$, and $\alpha(=\Delta\alpha)$ is similar to the case of a radiation-dominated period before inflation. Even better properties are obtained in this case, ($z \to 0$) $n_s - 1 \to -2(p+1)/p - 2(1+p) \cong 0.022$ for the spectral index in the case of $p = -10/9$) compared to $n_s - 1 \to -2(p+1)/p - 2p \cong 2.022$ in the radiation-dominated case, and a constant value of 1 for $|C(k)|^2$ in the $z \to \infty$ limit compared to oscillation around $2/\sqrt{3}$ in the radiation-dominated case.

**5.3. Matching conditions**

One of two matching conditions for the scalar perturbations was used in sections 5.1 and 5.2. The first is a matching condition in which the gauge potential and its first $\eta$-derivative are continuous at the transition point (Condition A). This matching condition allows the initial condition of pre-inflation to be decided rationally, that is, in the limit $\eta \to -\infty$, the field $u_k(\eta)$ approaches plane waves. The second matching condition is that of Deruelle and Mukhanov [9] for cosmological perturbations, which requires that the transition occurs on a hyper-surface of constant energy (Condition B).

The quantity $|C(k)|$, the spectral index and the running spectral index are calculated here for Deruelle's matching condition, which dictates that $\Phi$ and $\zeta$ (**R**) are continuous at the transition time $\eta_2$, and the difference between the two matching conditions is investigated. The parameters $\Phi$ and $\zeta$ can be written as follows [17,18].

$$\Phi = (-\boldsymbol{H}^2 + \boldsymbol{H}')\,(\frac{u}{Z})'/(c_s^2 k^2 \boldsymbol{H}), \tag{38}$$

$$\zeta = (\boldsymbol{H}\Phi' - \Phi\boldsymbol{H}' + 2\boldsymbol{H}^2\Phi)/(\boldsymbol{H}^2 - \boldsymbol{H}'), \tag{39}$$

where $Z$ is written as $a(\boldsymbol{H}^2 - \boldsymbol{H}')^{1/2}/(c_s \sqrt{4\pi G}\,\boldsymbol{H})$. As $\Phi$ and $\zeta$ can be written in terms of $u_k(\eta)$, the coefficients $c_1$ and $c_2$ are obtained as follows: In the period of pre-inflation, the mode function $u_k(\eta)$ is derived from equation (5), and $\Phi_P$ and $\zeta_P$ can be obtained using the relations (38) and (39). On the other hand, in inflation, $u_k(\eta)$ is expressed as $c_1 f_k^I(\eta) + c_2 f_k^{I*}(\eta)$, and this is used to calculate $\Phi_I$ and $\zeta_I$. From the relations $\Phi_P(\eta_2) =$ $\Phi_I(\eta_2)$ and $\zeta_P(\eta_2) = \zeta_I(\eta_2)$, the coefficients $c_1$ and $c_2$ can be fixed. Here, the matching condition of $\Phi$ and $\zeta$ ($R$) continuous at the transition point is adopted. However, the matching condition of Deruelle and Mukhanov must be written such that $\Phi$ and $\zeta + k^2\Phi/(3(H^2-H'))$ are continuous at the transition point. In the present case ($k\eta_2 = -z$), the value of $\zeta$ becomes smaller than that of the $k^2$ term, causing the latter to dominate at $|z| \gg 1$. However, the coefficients $c_1$ and $c_2$ cannot be fixed using $\Phi$ and the $k^2$ term. Thus, the matching condition of $\Phi$ and $\zeta$ ($R$) continuous at the transition point is adopted for the present treatment. As the calculation of the coefficients $c_1$ and $c_2$ is similar to that in previous sections, only the results are given here.

In the case of the radiation-dominated period before inflation, $|C(k)|$ can be written from equation (13) as

$$C(k) = \frac{\sqrt{\pi}}{4 \cdot 3^{3/4} \sqrt{p(p+1)}\sqrt{z}} e^{iz/\sqrt{3}p} \{(\sqrt{3}+4\sqrt{3}\,p^2 - p(\sqrt{3}+6iz))$$

$$\times (H^{(1)}_{-p+1/2}(z) + H^{(2)}_{-p+1/2}(z)) - \sqrt{3}\,(1+p)z(H^{(1)}_{-p+3/2}(z) + H^{(2)}_{-p+3/2}(z))\}. \qquad (40)$$

For $z \to 0$,

$$|C(k)|^2 \cong \frac{2^{-3+2p}\pi(-1+p-4p^2)^2 z^{-2p}}{\sqrt{3}p(1+p)(\Gamma(\frac{3}{2}-p))^2} + \frac{2^{-3+2p}\pi(3-2p-21p^2+24p^3+16p^4)\,z^{-2p+2}}{\sqrt{3}p(1+p)(-3+2p)(\Gamma(\frac{3}{2}-p))^2}.$$

$$(41)$$

For $z \to \infty$,

$$|C(k)|^2 \cong \frac{(1+2p+13p^2)+(1+2p-11p^2)\cos\theta}{4\sqrt{3}p(1+p)} + \frac{(-1+9p-3p^2+11p^3)\sin\theta}{4\sqrt{3}(1+p)z}$$

$$+\frac{1+3p-33p^2+61p^3+(1+2p+11p^2-53p^3-4p^4+11p^5)\cos\theta}{8\sqrt{3}(1+p)z^2} \tag{42}$$

where $\theta = p\pi + 2z$. The quantity $|C(k)|^2$ is plotted as a function of $z$ in figure 7 for $p=-10/9$. $|C(k)|^2$ becomes zero in the $z\to 0$ limit, and oscillates around $(p+1)/(2\sqrt{3}p) \leq |C(k)|^2 \leq (2\sqrt{3}p)/(p+1)$ in the $z\to\infty$ limit, the latter corresponding to a numerical range of $0.029 \leq |C(k)|^2 \leq 34.6$ for $p=-10/9$. Comparing figure 1 with figure 7, some differences between the matching conditions are apparent, such as the enhancement of $|C(k)|^2$ from large to small scales.

For the spectral indices of the contribution of pre-inflation, $\Delta n$, the asymptotic form of $\Delta n$ is obtained from equations (41) and (42) as $\Delta n \to -2p$ for $z \to 0$ and $\Delta n \to 0$ for $z \to \infty$. From equation (19), $z\to 0$, $n_s - 1 \to -2p - 2(p+1)/p$ and $z\to\infty$, $n_s - 1 \to -2(p+1)/p$. This asymptotic value is the same as that in section 5.1. In detail, in the case of $z \to 0$, using equation (41), the spectral index and the running spectral index can be calculated as follows.

$$\Delta n \cong -2p + \frac{(6-4p-42p^2+48p^3+32p^4)z^2}{(-3+2p)(-1+p-4p^2)^2}, \tag{43}$$

$$\Delta\alpha \cong \frac{2(6-4p-42p^2+48p^3+32p^4)z^2}{(-3+2p)(-1+p-4p^2)^2} \ . \tag{44}$$

For the same reason as discussed above, $n_s - 1$ is only shown for the range $0 \leq z \leq 3$ in figure 8.

A comparison of the two matching conditions reveals that while the behavior of $|C(k)|^2$ is similar, the values differ, but $\Delta n$ has the same value for both matching conditions concerning the leading order.

For the case of the scalar-matter-dominated period before inflation, $C(k)$ can be written from equation (13) as

$$C(k) = \frac{-\sqrt{\pi}}{16\sqrt{3p(p+1)}\sqrt{z^3}} \, i e^{2iz/p}$$

$$\{(p^3 + p^2(-4-2iz) - 8iz + p(4+8iz-12z^2))(H^{(1)}_{-p+1/2}(z) + H^{(2)}_{-p+1/2}(z))$$

$$- 4(1+p)(p-2iz)z(H^{(1)}_{-p+3/2}(z) + H^{(2)}_{-p+3/2}(z))\}. \tag{45}$$

In the case of $z \to 0$,

$$|C(k)|^2 \cong \frac{2^{-7+2p}\pi p(-2+p)^4 z^{-2-2p}}{3(1+p)(\Gamma(\frac{3}{2}-p))^2} + \frac{2^{-7+2p}\pi(-2+p)^3(24-28p-34p^2+p^3)\,z^{-2p}}{3p(1+p)(-3+2p)(\Gamma(\frac{3}{2}-p))^2}$$

$$\tag{46}$$

In the case of $z \to \infty$,

$$|C(k)|^2 \cong \frac{(4+8p+13p^2)+(4+8p-5p^2)\cos\theta}{12p(1+p)} + \frac{(-4-9p^2+5p^3)\sin\theta}{12(1+p)z}$$

$$+ \frac{(-2+p)(-4-16p+9p^2+(-4-12p-15p^2+10p^4)\cos\theta)}{48(1+p)z^2} \quad (47)$$

where $\theta = p\pi + 2z$. $|C(k)|^2$ is plotted as a function of $z$ in figure 9 for $p = -10/9$. $|C(k)|^2$ becomes zero in the $z \to 0$ limit, and oscillates around $2(p+1)/3p \leq |C(k)|^2 \leq 3p/(2(p+1))$ in the $z \to 0$ limit, the latter corresponding to a numerical range of $0.067 \leq |C(k)|^2 \leq 15$ for $p = -10/9$. Comparing figures 4 and 9, $|C(k)|^2$ is large and exhibits oscillation in figure 9, while the value is constant at 1 in figure 4.

Obtaining the asymptotic form of $\Delta n$ from equations (46) and (47): $\Delta n \to -2-2p$ for $z \to 0$ and $\Delta n \to 0$ for $z \to \infty$, from equation (19) $z \to 0$, $n_s - 1 \to -2 - 2p - 2(p+1)/p$ and $z \to \infty$, $n_s - 1 \to -2(p+1)/p$. This asymptotic value is the same as that in section 5.2. In detail, in the case of $z \to 0$, using equation (34), the spectral index and running spectral index are calculated as follows.

$$\Delta n \cong -2(1+p) + \frac{2(24-28p-34p^2+p^3)z^2}{p^2(6-7p+2p^2)}, \quad (48)$$

$$\Delta\alpha \cong \frac{4(24-28p-34p^2+p^3)z^2}{p^2(6-7p+2p^2)}. \quad (49)$$

Because of the same reason in section 5.1 and 5.2, $n_s - 1$ is shown only for the range $0 \leq z \leq 3$ in figure 10.

A comparison of the two matching conditions shows that while the behavior of $|C(k)|^2$ is similar in these two matching conditions with respect to a scalar-matter-dominated period before inflation, the values differ, but $\Delta n$ is the same for both matching conditions concerning the leading order.

Considering all four cases, the two models and two matching conditions, there are no large differences regarding the behavior of $|C(k)|^2$, the spectral index or the running spectral index. Thus, if the length of inflation is close to 60 *e*-folds, one of these scenarios may explain the WMAP data.

## 6. Discussion and summary

Two pre-inflation models were considered in this study, describing a radiation-dominated period or a scalar-matter-dominated period before inflation. The effect of pre-inflation was described by the factor $|C(k)|$, allowing the familiar formation of the derived power spectrum of curvature perturbations to be simply multiplied by this factor. Considering in addition two matching conditions, $|C(k)|$, the spectral index and the running spectral index were calculated, and the characteristics of each of the scenarios were compared. The properties were in fact found to be relatively constant across all four scenarios. The power spectra was found to decrease on super large scales, and oscillate from large scale to small scales. The spectral index was found to be $n_s - 1 > 0$ on large (super) scales, slowly oscillating and approaching $n_s - 1 \cong -2(p+1)/p$ at smaller scales. This implies $n_s - 1 < 0$ is possible.

These properties were shown to be similar to the four properties of WMAP data considered here. However, figures 2, 5, 8 and 10 revealed fairly large values of $n_s - 1$. As is well known, it appears difficult to derive the four properties of WMAP data using simple inflation models [2]. As the present models are very simple and involve only two parameters, *p* and the length of inflation, these models seem to be quite probable. However, we consider that it is still premature to propose the better of the models.

It should also be noted that the WMAP data does not express the spectral running index in detail, and there remains some debate regarding the decrease of the power spectrum on large scales [5]. On the other hand, it seems difficult that we completely fix the power spectrum at near the transition of inflation by simple our calculation. Here, the possibility has been shown that the present pre-inflation models explain the WMAP data. If these pre-inflation models are indeed correct, this will give the valuable result that the length of inflation must be close to 60 *e*-folds based on the WMAP data. Future analyses can be expected to provide a clearer answer as the precision of WMAP data increases.

**Acknowledgments**

The author would like thank the staff of Osaka Electro-Communication University for valuable discussions.

\* On leave from Osaka Electro-Communication Junior College

References

[1] Bennett C L et al. 2003 Astrophys. J., Suppl. Ser. **146** 1 astro-ph/0302207; Spergel D N et al. 2003 Astrophys. J., Suppl. Ser. **146**, 175 astro-ph/0302209; Peiris H V et al. 2003 Astrophys. J., Suppl. Ser. **146** 213 astro-ph/030225

[2] Chung D J H, Shiu G, and Trodden M 2003 Phys. Rev. **D68** 063501; Leach S M, Liddle A R 2003 Phys. Rev. **D68** 123508

[3] Kawasaki M, Yamaguchi M, and Yokoyama J 2003 Phys. Rev. **D68** 023508; Feng B and Zhang X 2003 Phys. Lett. B **570** 145; Contaldi C R, Peloso M, Kofman L and Linde A 2003 JCAP **0307** 002 astro-ph/0303636

[4] Tegmark M, de Costa-Oliviera A, and Hamilton A 2003 Phys. Rev. **D68** 123523 astro-ph/0302496; Luminet J-P, Weeks J R, Riazuelo A, Lehoucq R and Uzan, J -P 2003 Nature **425** 593; Uzan J P, Riazuelo A, Lehoucq R, and Weeks J astro-ph/0302580; Efstathiou G 2003 Mon. Not. R. Astron. Soc. **343** L95; Uzan J P, Kirchner U and Ellis G F R 2003 Mon. Not. R. Astron. Soc. **344** L65

[5] Bastero-Gil M, Freese K, Mersini-Houghton L 2003 Phys. Rev. **D68** 123514

[6] Elgaroy O, and Hannestad S 2003 Phys. Rev. **D68** 123513

[7] Hirai S. 2003 Class. Quantum Grav. **20** 1673 hep-th/0212040; Hirai S 2003 hep-th/0307237

[8] Grishchuk L P and Sidorov Y V 1989 Class. Quantum Grav. **6** L161; Grishchuk L P, Haus H A and Bergman K 1992 Phys. Rev. **D46** 1440; Giovannini, M 2000 Phys. Rev. **D61** 087306; Allen B, Flanagan E E, Papa M A 2000 Phys. Rev. **D61** 024024; Hirai S

2000 Prog. Theor. Phys. **103** 1161; Kiefer C, Polarski D, and Starobinsky A A 2000 Phys. Rev. **D62** 043518

[9]  Deruelle N and Mukhanov V F 1995 Phys. Rev. **D52** 5549

[10] Durrer R and Vernizzi F 2002 Phys. Rev. **D66** 83503

[11] Lidsey J E, LiddleA R, Kolb E W, Copeland E J, Barreiro T, and Abney M 1997 Rev. Mod. Phys. **69** 373

[12] Hwang J 1998 Class. Quantum Grav. **15** 1387 *ibid*.1401

[13] Bardeen J M 1980 Phys. Rev. **D22** 1882; Kodama H and Sasaki M 1984 Prog. Theor. Phys. Suppl. **78** 1

[14] Mukhanov V F, Feldman H A and Brandenberger R H 1992 Phys. Rep. **215** 203

[15] Handbook of Mathematical Functions, edited by Abramaswits M and Stegun I (Dover, NewYork, 1970)

[16] Liddle A R and Lyth D H 1993 Phys. Rep .**231** 1

[17] Albrecht A, Ferreira P, Joyce M, and Prokopec T 1994 Phys. Rev. **D50** 4807

[18] Mukhanov V F, Feldman H A, and Brandenberger R H 1992 Phys. Rep. **215** 203

[19] MonizP V, Mourao J M, and Sa P M 1993 Class. Quantum Grav. **10** 517

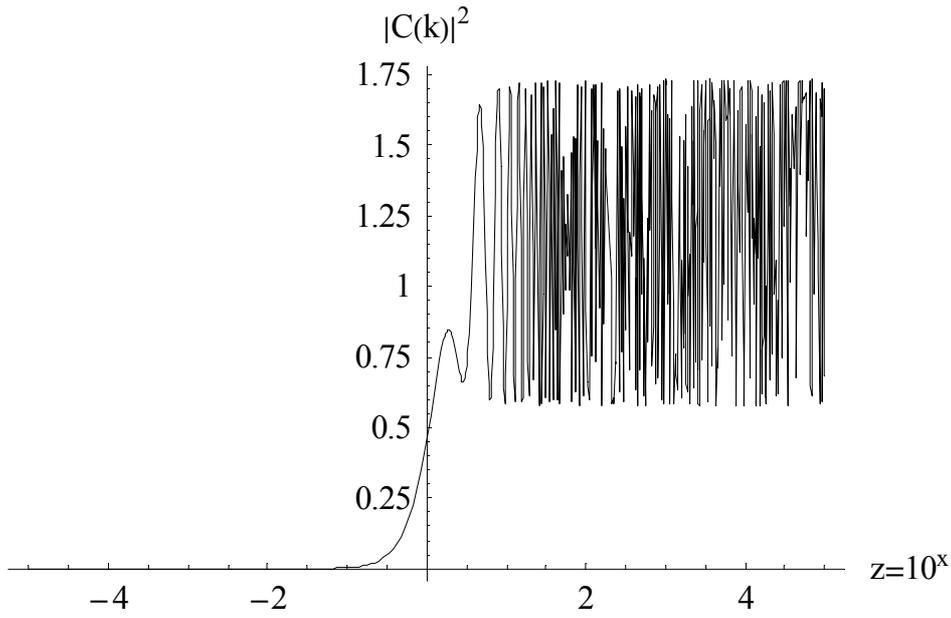

Figure 1. Factor $|C(k)|^2$ as a function of $z$ $(= -k\eta_2)$ for $10^{-5} \leq z \leq 10^5$ in the case of a radiation-dominated period before inflation under matching condition A ($p = -10/9$)

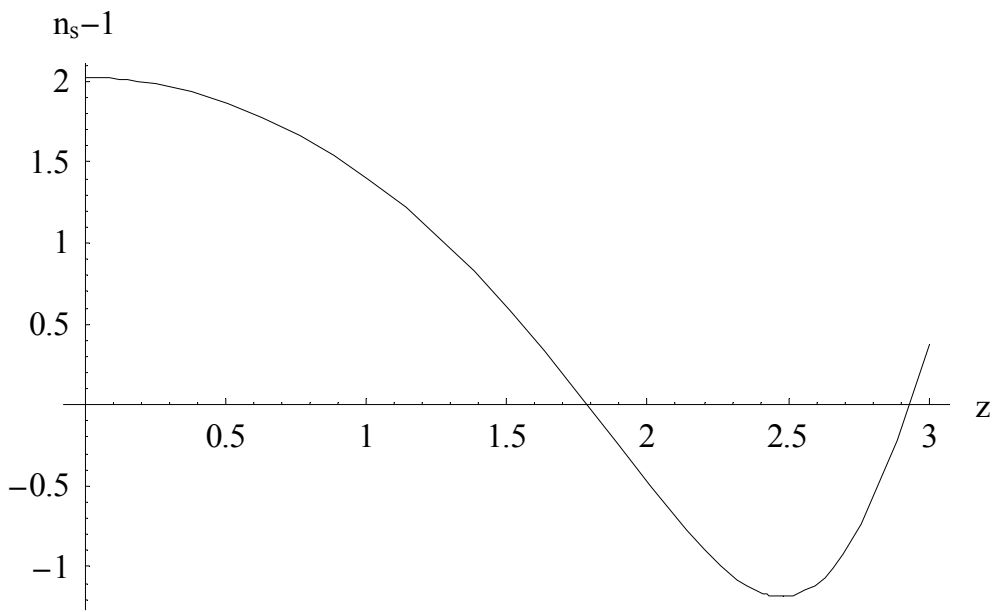

Figure 2. Spectral index $n_s - 1$ as a function of $z\ (= -k\eta_2)$ for $0 \leq z \leq 3$ in the case of a radiation-dominated period before inflation under matching condition A ($p = -10/9$)

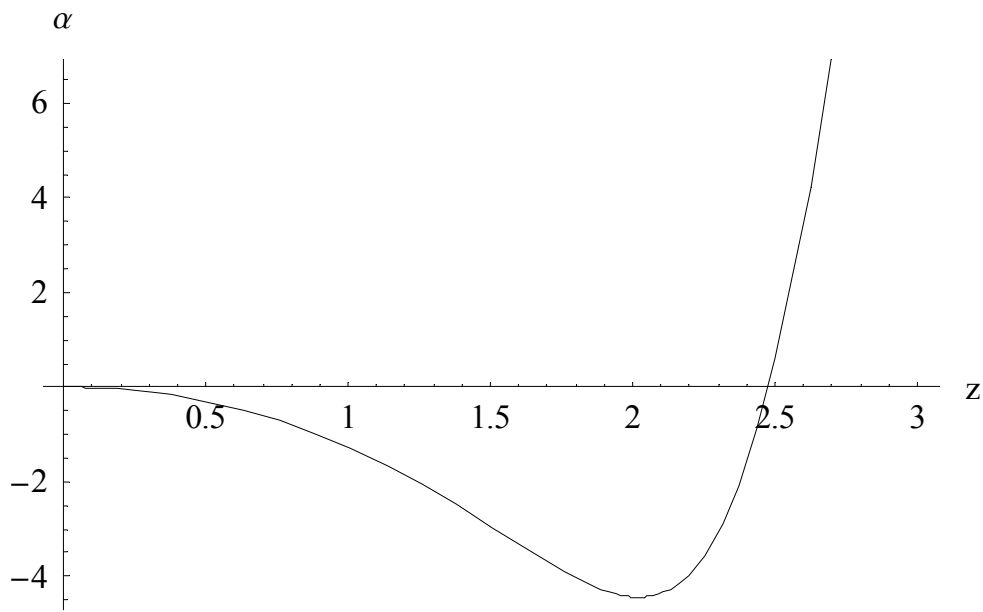

Figure 3. Running spectral index $\alpha$ as a function of $z\ (= -k\eta_2)$ for $0 \leq z \leq 3$ in the case of a radiation-dominated period before inflation under matching condition A ($p = -10/9$)

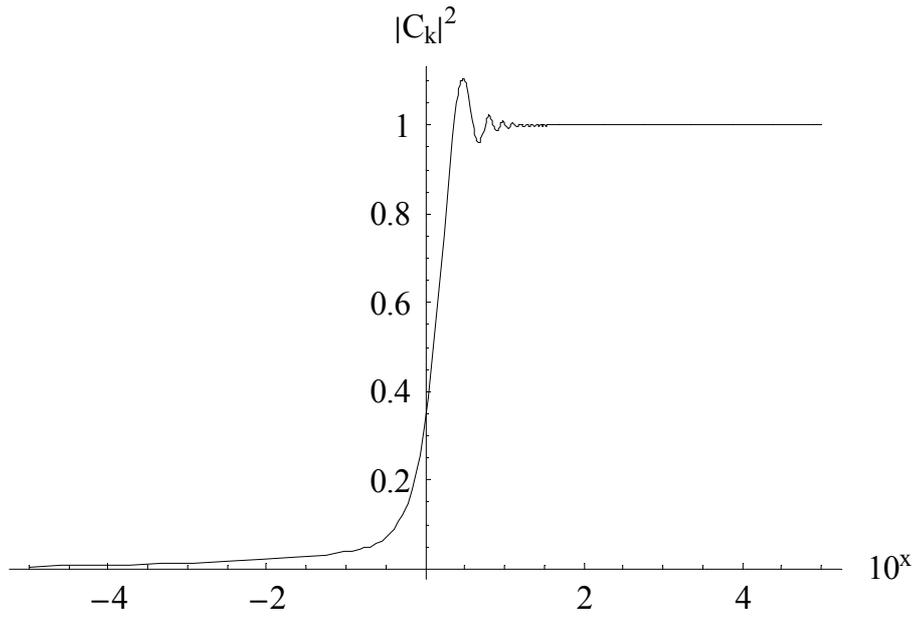

Figure 4. Factor $|C(k)|^2$ as a function of $z$ $(= -k\eta_2)$ for $10^{-5} \leq z \leq 10^5$ in the case of a scalar-matter-dominated period before inflation under matching condition A ($p = -10/9$)

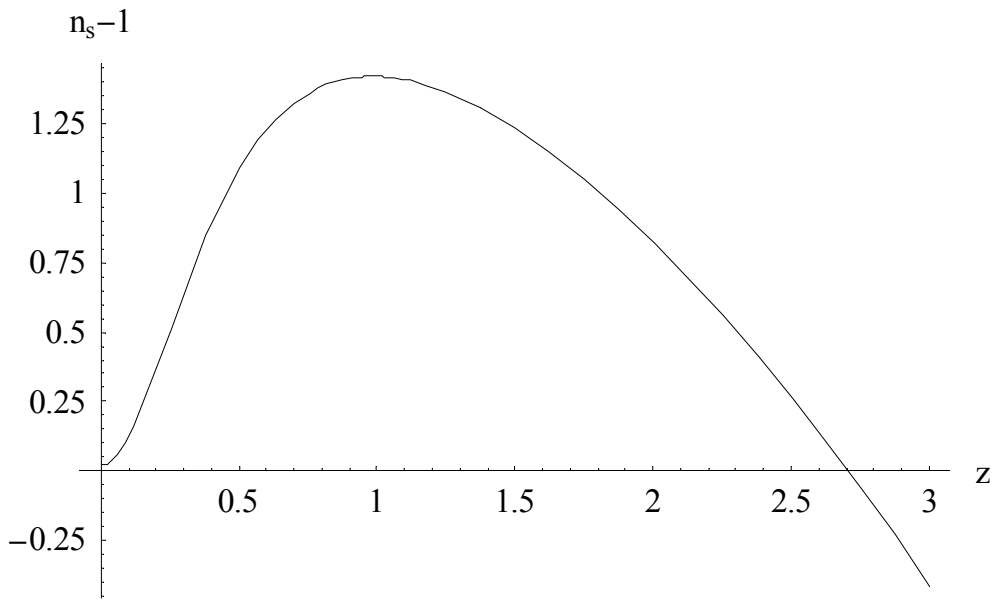

Figure 5. Spectral index $n_s - 1$ as a function of $z$ $(= -k\eta_2)$ for $0 \leq z \leq 3$ in the case of a scalar-matter-dominated period before inflation under matching condition A ($p = -10/9$)

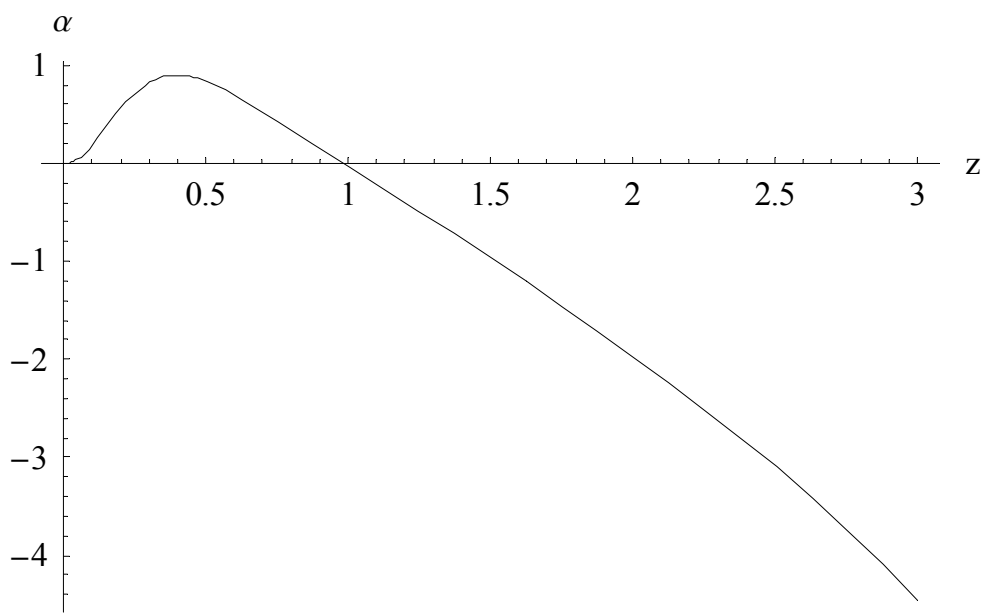

Figure 6. Running spectral index $\alpha$ as a function of $z$ $(= -k\eta_2)$ for $0 \leq z \leq 3$ in the case of a scalar-matter-dominated period before inflation under matching condition A ($p = -10/9$)

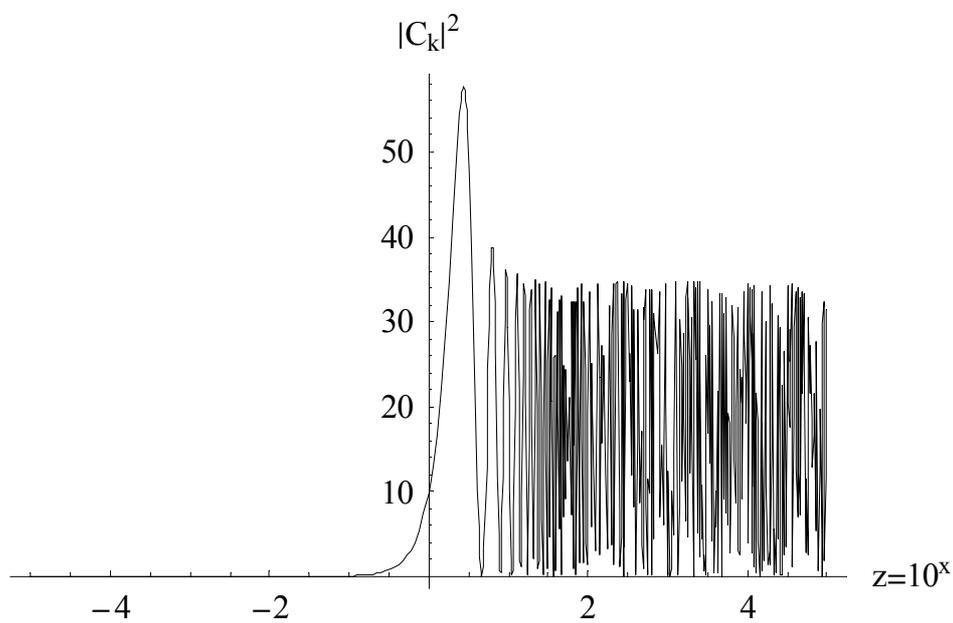

Figure 7. Factor $|C(k)|^2$ as a function of $z$ $(=-k\eta_2)$ for $10^{-5} \leq z \leq 10^5$ in the case of a radiation-dominated period before inflation under matching condition B $(p=-10/9)$

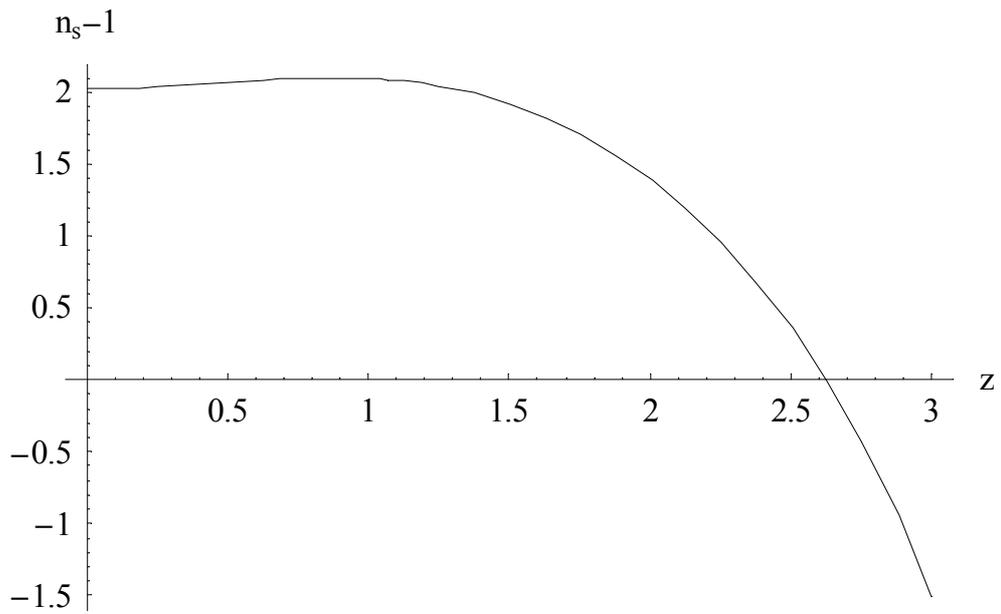

Figure 8. Spectral index $n_s -1$ as a function of $z$ $(=-k\eta_2)$ for $0 \leq z \leq 3$ in the case of a radiation-dominated period before inflation under matching condition B $(p=-10/9)$

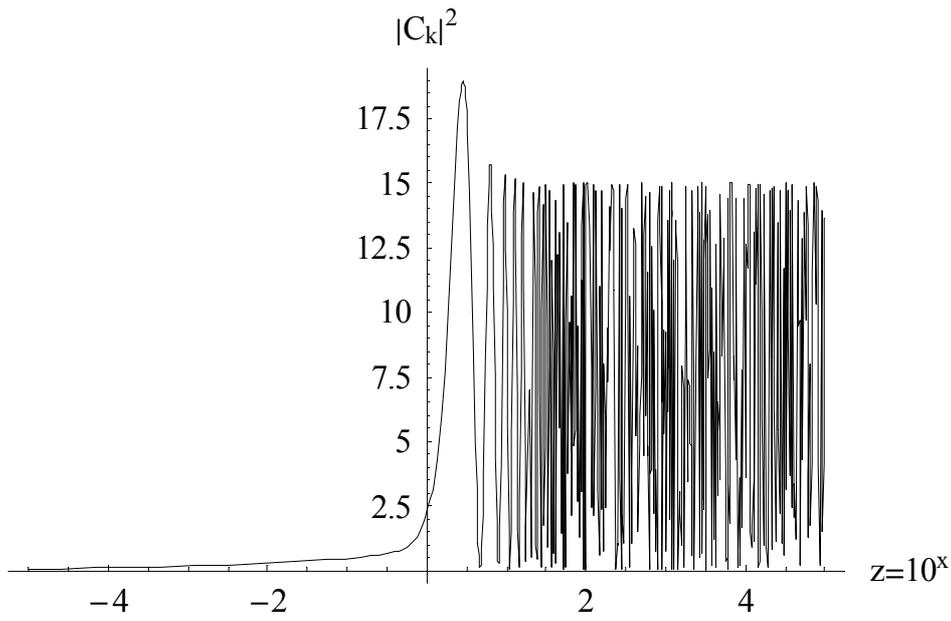

Figure 9. Factor $|C(k)|^2$ as a function of $z$ $(= -k\eta_2)$ for $10^{-5} \leq z \leq 10^5$ in the case of a scalar-matter-dominated period before inflation under matching condition B ($p = -10/9$)

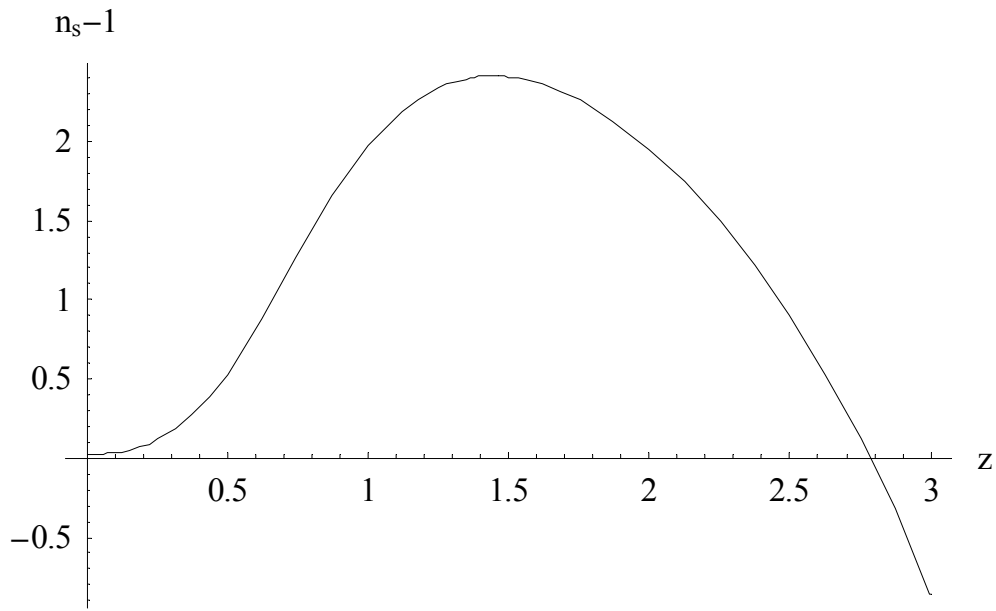

Figure 10. Spectral index $n_s - 1$ as a function of $z$ $(=-k\eta_2)$ for $0 \leq z \leq 3$ in the case of a scalar-matter-dominated period before inflation under matching condition B ($p = -10/9$)

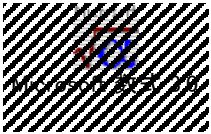